\documentclass[11pt,titlepage]{article}

\usepackage{graphicx}
\graphicspath{{./}}
\usepackage{amssymb,amsfonts,amsmath}
\usepackage{rotating}
\usepackage{cite}

\newcommand{\citep}[1]{\cite{#1}}
\newcommand{\pcitep}[2]{(#1~\citep{#2})}

\newcommand*{\abbr}[1]{#1}
\newcommand*{\dfn}[1]{\emph{#1}}
\newcommand*{\tech}[1]{{\sc #1}}

\newcommand*{\HSR}{heat-shock response}
\newcommand*{\HSP}{heat-shock proteins}
\newcommand*{\HSF}{heat-shock factor}
\newcommand*{\HSE}{heat-shock element}
\newcommand*{\PMC}{probabilistic model checking}

\newcommand*{\CTMC}{continuous-time Markov process}
\newcommand*{\ODE}{ordinary differential equations}
\newcommand*{\CME}{chemical master equation}

\newcommand*{\poc}{proof of concept}
\newcommand*{\cs}{case study}

\newcommand{\invit}{\textit{in~vitro}}

\newcommand*{\Ma}{Mass action}
\newcommand*{\ma}{\MakeLowercase{\Ma}}

\newcommand{\cdeg}{\ensuremath{^{\circ}\text{C}}}
\newcommand*{\fu}[1]{\ensuremath{\mathrm{#1}}} 
\newcommand*{\molar}{\ensuremath{\fu{M}}} 
\newcommand{\tu}{\ensuremath{\fu{min}}} 
\newcommand{\cu}{\ensuremath{\fu{a.s.}\molar}}
\newcommand*{\xu}[1]{\ensuremath{\ \fu{#1}}} 

\newcommand{\thspsi}{Hsp70}

\newcommand{\tS}{substrate}
\newcommand{\thsp}{HSP}
\newcommand{\thsf}{HSF}
\newcommand{\tmrna}{\thsp\ mRNA}

\newcommand{\thsps}{\thsp:\tS}
\newcommand{\thsfA}{\thsf$_3$}
\newcommand{\thse}{HSE}
\newcommand{\thseO}{\thse:\thsfA}
\newcommand{\tP}{native protein}

\newcommand{\xfont}[1]{\textrm{#1}}
\newcommand{\xS}{\ensuremath{\xfont{S}}}
\newcommand{\xhsp}{\ensuremath{\xfont{HSP}}}
\newcommand{\xhsf}{\ensuremath{\xfont{HSF}}}
\newcommand{\xmrna}{\ensuremath{\xfont{mRNA}}}
\newcommand{\xhsphsf}{\ensuremath{\xhsp\text{:}\xhsf}}
\newcommand{\xhsps}{\ensuremath{\xhsp\text{:}\xS}}
\newcommand{\xhsfA}{\ensuremath{\xhsf_3}}
\newcommand{\xhse}{\ensuremath{\xfont{HSE}}}
\newcommand{\xhseO}{\ensuremath{\xhse\text{:}\xhsfA}}
\newcommand{\xP}{\ensuremath{\xfont{P}}}

\newcommand{\xT}{\ensuremath{\text{T}}}

\newcommand{\kr}[2]{\ensuremath{\xfont{#1}_{#2}}}

\RequirePackage{xargs}
\RequirePackage{mhchem}
\newcommandx*{\reacts}[1][1=]{\ensuremath{%
\ce{->[#1]}
}}
\newcommandx*{\reactsf}[1][1=]{\ensuremath{%
\ce{->[~~#1~~]}
}}
\newcommandx*{\rreacts}[2][1=,2=]{\ensuremath{%
\ce{<=>[#1][#2]}
}}
\newcommandx*{\rreactsf}[2][1=,2=]{\ensuremath{%
\ce{<=>>[#1][#2]}
}}
\newcommandx*{\rreactsb}[2][1=,2=]{\ensuremath{%
\ce{<<=>[#1][#2]}
}}

\newcommandx*{\fofx}[2]{\ensuremath{#1\left(#2\right)}}
\newcommand*{\symTime}{\ensuremath{t}}
\newcommandx*{\fAt}[2][2=\symTime]{\fofx{#1}{#2}}
\newcommandx*{\hfrac}[4][1=.,4=.]{\left#1{#2}\middle/{#3}\right#4}
\newcommand*{\newop}[1]{\ensuremath{\operatorname{#1}}}
\newcommand*{\e}{\newop{e}}
\newcommand*{\opD}{\newop{\mathrm{d}}} 
\newcommandx*{\dx}[2][1=\opD]{\ensuremath{#1#2}}
\newcommandx*{\dt}[1][1=\opD]{\ensuremath{\dx[#1]{\symTime}}}
\newcommand*{\opMean}{\newop{\mathbb{E}}}

\newcommand*{\symSs}{\ensuremath{\pi}} 

\newcommand*{\symSpec}{\ensuremath{S}}
\newcommandx{\symSpecNr}{\ensuremath{\#}}
\newcommandx*{\specNr}[2][1=\symSpec,2=\symSpecNr]{\ensuremath{#2#1}}
\newcommandx*{\specConc}[1][1=\symSpec]{\ensuremath{\left[#1\right]}}

\newcommandx*{\specNrAt}[2][1=\specI,2=\symTime]{\fAt{\specNr[#1]}[#2]}
\newcommandx*{\specConcAt}[2][1=\specI,2=\symTime]{\fAt{\specConc[#1]}[#2]}

\newcommand{\dc}{\ensuremath{\mathcal{D}}}
\newcommand{\resp}{\ensuremath{\mathcal{R}}}
\newcommandx{\tgap}[1][1=\symTime]{\ensuremath{\Delta #1}}

\newcommand{\xI}{\ensuremath{\text{I}}}
\newcommand{\xImax}{\ensuremath{\xI_{100}}}
\newcommand{\tImax}{\ensuremath{\symTime_{100}}}

\newcommand{\tIhl}{\ensuremath{\symTime_{50}}}
\newcommand{\ki}{\kr{k}{\xI}}

\title{Modelling the efficacy of hyperthermia treatment}

\author{%
Miko{\l}aj Rybi{\'n}ski%
\thanks{
           Corresponding author. Current address:
           Department of Biosystems Science and Engineering,
           ETH Zurich,
           Mattenstrasse 26,
           4058 Basel, Switzerland,
           Tel.:~(004161)387-3385, Fax:~(004161)387-3994%
} \\
Institute of Informatics, \\
University of Warsaw, Warsaw, Poland \\
and \\
Mossakowski Medical Research Centre, \\
Polish Academy of Sciences, Warsaw, Poland
\and Zuzanna Szyma{\'n}ska \\
Interdisciplinary Centre for Mathematical and Computational Modelling, \\
University of Warsaw, Warsaw, Poland
\and S{\l}awomir Lasota \\
Institute of Informatics, \\
University of Warsaw, Warsaw, Poland
\and Anna Gambin \\
Institute of Informatics, \\
University of Warsaw, Warsaw, Poland \\
and \\
Mossakowski Medical Research Centre, \\
Polish Academy of Sciences, Warsaw, Poland%
}

\date{\today}

\begin{document}

\maketitle

\abstract{
Multimodal oncological strategies which combine chemotherapy or radiotherapy
with hyperthermia have a potential of improving the efficacy of the non-surgical
methods of cancer treatment.
Hyperthermia engages the heat-shock response mechanism (HSR), main component of
which are heat-shock proteins (HSP).
Cancer cells have already partially activated HSR, thereby,
hyperthermia may be more toxic to them relative to normal cells.
On the other hand, HSR triggers thermotolerance, i.e. hyperthermia treated cells
show an impairment in their susceptibility to a~subsequent heat-induced stress.
This poses questions about efficacy and optimal strategy of
the anti-cancer therapy combined with hyperthermia treatment.

To address these questions, we adapt our previous HSR model and propose its
stochastic extension.
We formalise the notion of a~HSP-induced thermotolerance. Next, we estimate
the intensity and the duration of the thermotolerance. Finally, we quantify the
effect of a~multimodal therapy based on hyperthermia and a~cytotoxic effect of
bortezomib, a~clinically approved proteasome inhibitor. Consequently, we
propose an optimal strategy for combining hyperthermia and proteasome
inhibition modalities.

In summary, by a~\poc\ mathematical analysis of HSR we are able to support the
common belief that the combination of cancer treatment strategies increases
therapy efficacy.

\vspace{2em}
\ma\ kinetics%
}

\clearpage

\section*{Introduction}

Most of the non-surgical methods of cancer treatment
(e.g.~chemotherapy and radiotherapy) are based on the principle of putting some
kind of stress on cancer cells to induce their death.
Unfortunately, in many cases the above methods fail.
The fact that \dfn{\HSP} (\abbr{HSP}) prevent apoptosis
induced by different modalities of cancer treatment explains how these proteins
could limit the application of such anti-cancer therapies~\citep{mayer2005}.
In order to improve the efficacy of these treatments, some effort is focused on
the multimodal oncological strategies which usually combine treatment of chemo-
or radiotherapy with hyperthermia.

\subsection*{Heat-shock response in cancer treatment}

\abbr{HSP} are a~group of highly conserved proteins involved
in many physiological and pathological cellular processes. They are so called
chaperones, as they protect proteins from stress and help new and
distorted proteins with folding into their proper
shape~\citep{georgopoulos1993}.
In principle, \abbr{HSP} synthesis increases under stress conditions.
Subsequently, upregulation of \abbr{HSP} increases cell survival and
stress-tolerance~\citep{parsell1993}.
Elevated expression of different members of \abbr{HSP} family has been detected
in several cases of tumour \pcitep{see, e.g.,}{barnes2001}. Despite its
importance, little is still known about how exactly \abbr{HSP} are involved in
different processes related to cancer development.
In this work we are interested in the heat-shock inducible isoform of
\HSP\ 70\xu{\fu{kDa}} (\thspsi).
For a~sake of clarity, we will denote \thspsi\ protein by \thsp, and use the
former only if context might be unclear.

Hyperthermia is a~therapeutic procedure used to raise the temperature
of a~whole body or a~region of the body affected by cancer.
Body tissues are, globally or locally, exposed to temperatures up to
45\cdeg~\citep{wust2002}.
Besides characteristics specific to cell type, the effectiveness of
hyperthermia depends on the temperature achieved during the treatment, as well
as on the length of the treatment~\citep{wust2002,hildebrandt2002}.
In general, moderate hyperthermia treatment, which maintains temperatures in
a~moderate 40--42\cdeg\ range for about an hour, does not damage most of normal
tissues and has acceptable adverse effects~\citep{wust2002,van_der_zee2002}.

Currently, hyperthermia effectiveness is under study in clinical trials,
including combination with other cancer therapies
\citep{wust2002,van_der_zee2002}.
A~synergistic interaction of radiotherapy and hyperthermia as well as some
cytotoxic drugs and hyperthermia has already been confirmed in experimental
studies~\citep{hildebrandt2002}.
In particular, Neznanov et al.~\citep{neznanov2011} demonstrated, \invit, that
induction of \dfn{\HSR} (\abbr{HSR}) by hyperthermia enhances the efficacy of
a~proteasome inhibitor called bortezomib~---~a~FDA-approved drug for treatment
of a~multiple myeloma and mantle cell lymphoma~\citep{molineaux2012}.
Basically, hyperthermia engages the \abbr{HSR} mechanism, main component of
which are the anti-apoptotic \abbr{HSP}.
Cancer cells have already partially activated \abbr{HSR}
because they are constitutively coping with higher level of misfolded protein
(mainly due to rapid rate of proliferation and specific intracellular
conditions of cancer cells). Therefore, in principle, sufficiently increased
level of misfolded proteins, as obtained by, e.g., hyperthermia, can not be
matched by cell's \abbr{HSR} capacity and, in effect, such enhanced proteotoxic
stress can be more toxic to them relative to normal cells \citep{neznanov2011}.

On the other hand, after a heat-shock, all cell types show an impairment in
their susceptibility to heat-induced cytotoxicity. This phenomenon, known as
\dfn{thermotolerance}, is triggered by \abbr{HSR} and it is, at least partially
based, on the upregulation of \abbr{HSP}~\citep{hildebrandt2002}.
Thermotolerance is, in principle, reversible and persists for usually between 24
and 48 hours~\citep{wust2002}. Due to this phenomenon the applicability of the
combined hyperthermia therapy may be, counter-intuitively, initially limited.
This naturally poses questions about the efficacy and about an optimal strategy
of the hyperthermia treatment.

\subsection*{Our results}

We formalise the notion of the \abbr{HSP}-induced thermotolerance i.e. the
memory of the \abbr{HSR} system of the previous temperature perturbation, or,
the system desensitisation with respect to the second consecutive heat-shock.
Using mathematical modelling we compute the intensity and the
duration of the thermotolerance. Finally, we give a~quantification of
an~effect of a combined therapy of hyperthermia and an bortezomib-induced
proteasome inhibition. Based on that, we propose an optimal strategy for
combination of heat-shock and the inhibitor. In principle, our results support
the common belief that the combination of aforementioned cancer treatment
strategies increases therapy efficacy.

\section*{Model}

The main purpose of this work is to contribute to the understanding of the
involvement of the  \abbr{HSR} mechanism in multimodal cancer therapies.
To this end, we use a~refined version of our previous deterministic
model~\citep{szymanska2009}.
Despite of its simplicity it provides a sound qualitative description of the
\abbr{HSR} mechanism.

\begin{figure}[!h]
\begin{center}
\includegraphics[width=.99\linewidth]{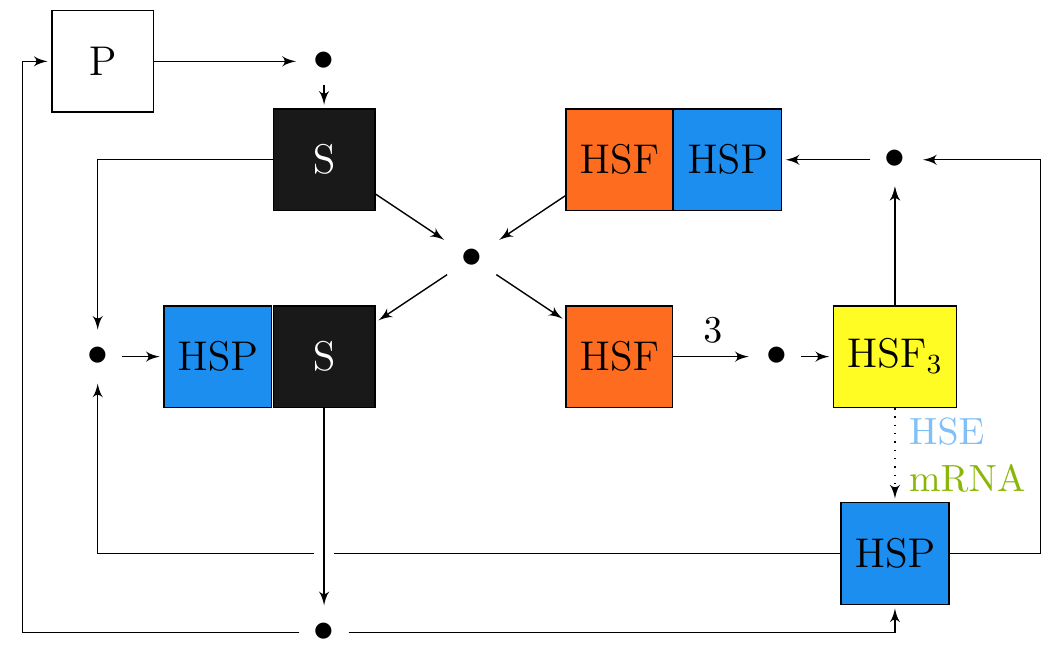}
\end{center}
\caption{%
}
\label{fig:hsr_scheme}%
\end{figure}

This model captures dynamics of synthesis of \thsp\ and its
interactions with key intracellular components of \abbr{HSR}, i.e.:
\abbr{HSP}; the \dfn{\HSF} (\abbr{HSF}) and its trimer,
which is a~\abbr{HSP} transcription factor; \abbr{HSP} \tS~---~mainly
denatured, misfolded {\tP}s; \abbr{HSP} gene~---~\dfn{\HSE}
(\abbr{HSE}); and \tmrna. Fig.~\ref{fig:hsr_scheme} depicts the overall
model scheme, and following reactions give the precise model structure:
\newcommand{\hs}{\hspace{2pt}}
\begin{align*}
\tag{r1}\label{eq:r1}
\xhsphsf + \xS &\ \ \ \rreactsf\ \ \  \xhsps + \xhsf,\\
\tag{r2}\label{eq:r2}
3\cdot\xhsf  &\ \ \ \ \ \reacts\ \ \ \  \xhsfA,\\
\tag{r3}\label{eq:r3}
\xhsfA + \xhse &\ \ \ \ \ \rreacts\ \ \ \hs \xhseO,\\
\tag{r4}\label{eq:r4}
\xhseO &\ \ \ \ \ \reacts\ \ \ \  \xhseO + \xmrna,\\
\tag{r5}\label{eq:r5}
\xhsp + \xhsfA &\ \ \ \ \ \reacts\ \ \ \ \hs \xhsphsf + 2\cdot\xhsf,\\
\tag{r6}\label{eq:r6}
\xhsp + \xS &\ \ \ \  \rreactsf\ \  \xhsps,\\
\tag{r7}\label{eq:r7}
\xhsp+\xhsf &\ \ \ \  \rreactsf\ \  \xhsphsf,\\
\tag{r8}\label{eq:r8}
\xhsp &\ \ \ \ \  \reacts\ \ \ \ \hs \emptyset,\\
\tag{r9}\label{eq:r9}
\xhsps &\ \ \ \ \  \reacts\ \ \ \hs \xhsp + \xP,\\
\tag{r10}\label{eq:r10}
\xP &\ \ \ \ \  \reacts[\xT]\ \ \ \  \xS,\\
\tag{r11}\label{eq:r11}
\xmrna &\ \ \ \ \  \reacts\ \ \ \ \  \xmrna + \xhsp,\\
\tag{r12}\label{eq:r12}
\xmrna &\ \ \ \ \  \reacts\ \ \ \ \  \emptyset.
\vspace{-2em}
\end{align*}
Four out of twelve reactions (Eqs~\ref{eq:r1}--\ref{eq:r12}) are reversible,
making it sixteen reactions in total;
\xT\ superscript over the reaction arrow
denotes temperature dependence. The proteins denaturation level dependence on
temperature is modelled by a~power-exponential function, analogously to
some of the previous \abbr{HSR} mathematical models
\citep{peper1998,szymanska2009,petre2011,mizera2010}. This type of functional
relation is based on an~experimental calorimetric enthalpy data
\citep{lepock1993}.

\begin{figure}
\begin{center}
\includegraphics[width=0.99\linewidth]{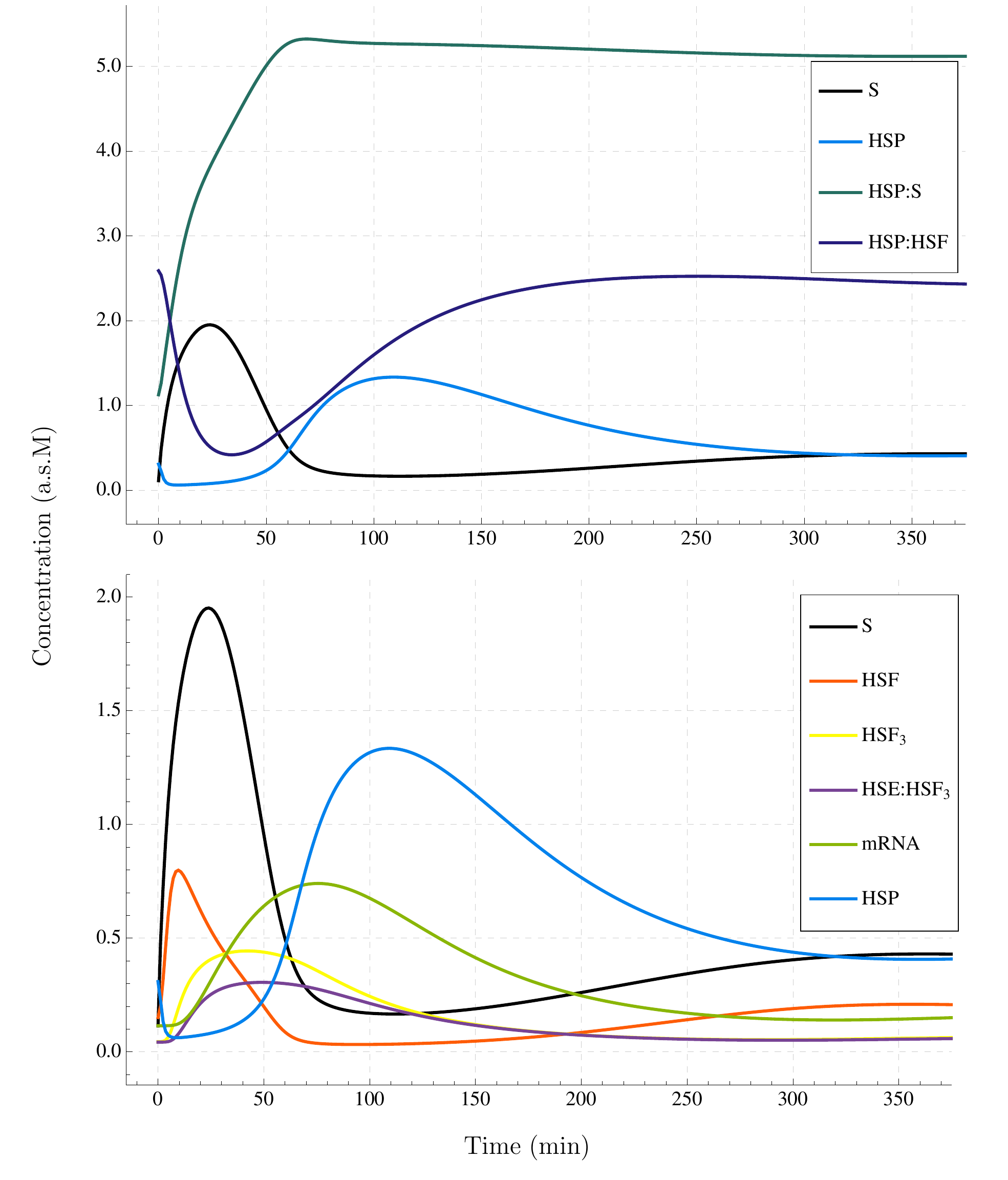}
\end{center}
\caption{%
}
\label{fig:hsr_activity}%
\end{figure}

Deterministic mathematical model follows purely \ma\ kinetics and it is
represented by the first order {\ODE} (\abbr{ODE}).
Fig.~\ref{fig:hsr_activity} depicts behaviour of this model,
which starts in the state of homeostasis, i.e. in a~steady state for
$\xT=37\cdeg$, in response to the immediate shift of the temperature
to~$\xT=42\cdeg$.
Amounts of species are arbitrarily scaled, each of them separately, to
obtain values of a~similar order of magnitude for each species (denoted \cu).
We calibrated this model with respect to the \thseO\ 42\cdeg\ experimental
data \citep{abravaya1991} (Fig.~\ref{fig:hsr_fit} in the Supporting Material).%

We additionally developed a~stochastic
counterpart of the deterministic model, represented by \dfn{\CME} (\abbr{CME})
or, equivalently, \dfn{\CTMC} (\abbr{CTMC}), which we then analysed using the
\dfn{\PMC} technique (\abbr{PMC}).
To ensure the feasibility of this approach we used approximate \abbr{PMC}
techniques, implemented recently in the \tech{PRISM}
tool~\citep{kwiatkowska2011}.

We found deterministic approach to be a~valid approximation of stochastic model,
i.e., both variants presented very similar results with respect to the
stochastic mean. However, almost half of the modelled species exhibit
a~significant noise level in the stochastic model. See
Text~\ref{txt:hsr_stoch} in the Supporting Material, Sec.~1 for details.
Both deterministic and stochastic models are additionally available in the
Supporting Material, respectively, as a~XML File~\ref{fil:sbml} in the
\tech{SBML} format~\citep{hucka2003}, as well as a~text File~\ref{fil:prism} in
the \tech{PRISM} model format~\citep{kwiatkowska2011}.

\section*{Results}

\subsection*{Quantification of the thermotolerance phenomenon}
Thermotolerance can be described as a~desensitisation with respect to
a~consecutive heat-shock, compared to the response to the first heat-shock. In
other words, thermotolerance represents a~memory of the system about the first
two, ``on'' and ``off'' temperature perturbations, leading to a~decreased
response to the subsequent ``on'' perturbation.
In case of the \abbr{HSR} system, its memory is created by a~propagating shift
in species activity and the feedback loop of the biochemical network
(cf. Fig.~\ref{fig:hsr_activity}).

\begin{figure}
\begin{center}
\includegraphics[width=0.99\linewidth]{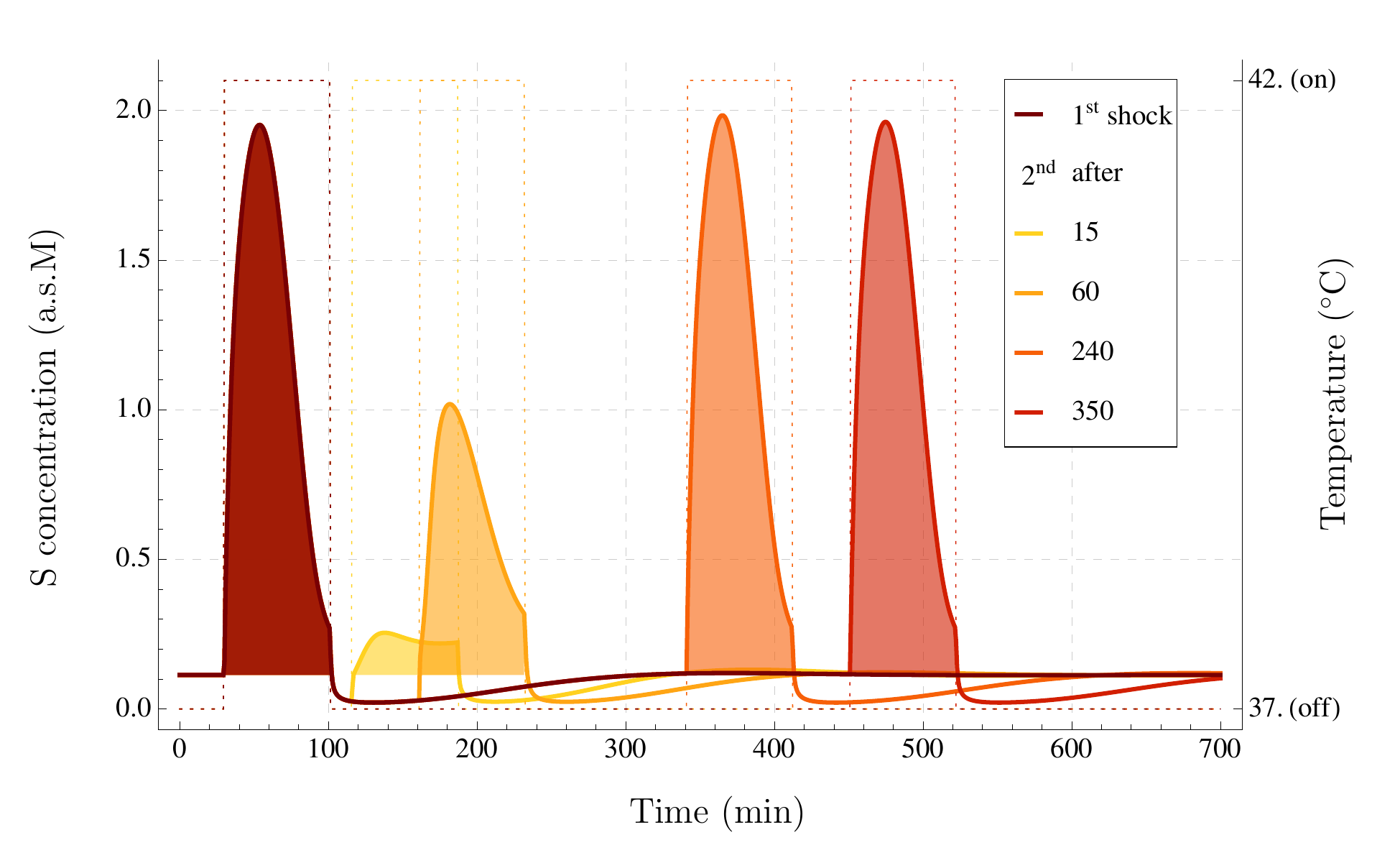}
\end{center}
\caption{%
}
\label{fig:hsr_desens}%
\end{figure}

Fig.~\ref{fig:hsr_desens} depicts the thermotolerance phenomena in the
deterministic \abbr{HSR} model for the immediate 42\cdeg\ heat-shock. Duration
and strength of the memory of the first temperature perturbation can be
accurately tracked by the activity of \thsp, level of which is negatively
correlated with the strength of the response (cf.~Fig.~\ref{fig:hsr_desens_shift}
in the Supporting Material).

In the stochastic model we introduce approximate perturbations as an
independent, $k$-level Poisson process (see Text~\ref{txt:hsr_stoch} in the
Supporting Material, Sec.~3 for details).
This allows to stay within the same mathematical model, i.e., \abbr{CTMC}, and
seamlessly perform stochastic simulations and model checking.

We define the notion of the \abbr{HSP}-induced thermotolerance during $n$-th
heat-shock ($n>1$) as the \dfn{desensitisation coefficient}:
\begin{equation}\label{eq:desens_coeff}
\dc_{n}=1-\frac{\resp_n}{\resp_{1}},
\end{equation}
where $n$-th response $\resp_n$ is defined as:
\begin{equation}\label{eq:response}
\resp_n=\max_{t_n\leq t<t_{n+1}}
\left\{\specNrAt[\symSpec]-\specNr^{*}\right\},
\end{equation}
where
$\specNr^{*}=\fofx{\opMean_\symSs}{\specNr}$
is a~mean value of a~species \symSpec\ amount in a~steady state $\symSs$;
$t_n$ is a~$n$-th heat-shock start time (we assume $t_{n+1}=\infty$ if not
specified otherwise); and the first response, by assumption, satisfies
$\resp_{1}>0$.
Such response measure represents the toxicity of the heat-shock: the higher the
response the more likely the cell will die.
For the deterministic model the species amount is simply a~scaled value of
\abbr{ODE} variable, corresponding to the mean value of a~stochastic process
random variable.

\begin{figure}
\begin{center}
\includegraphics[width=0.99\linewidth]{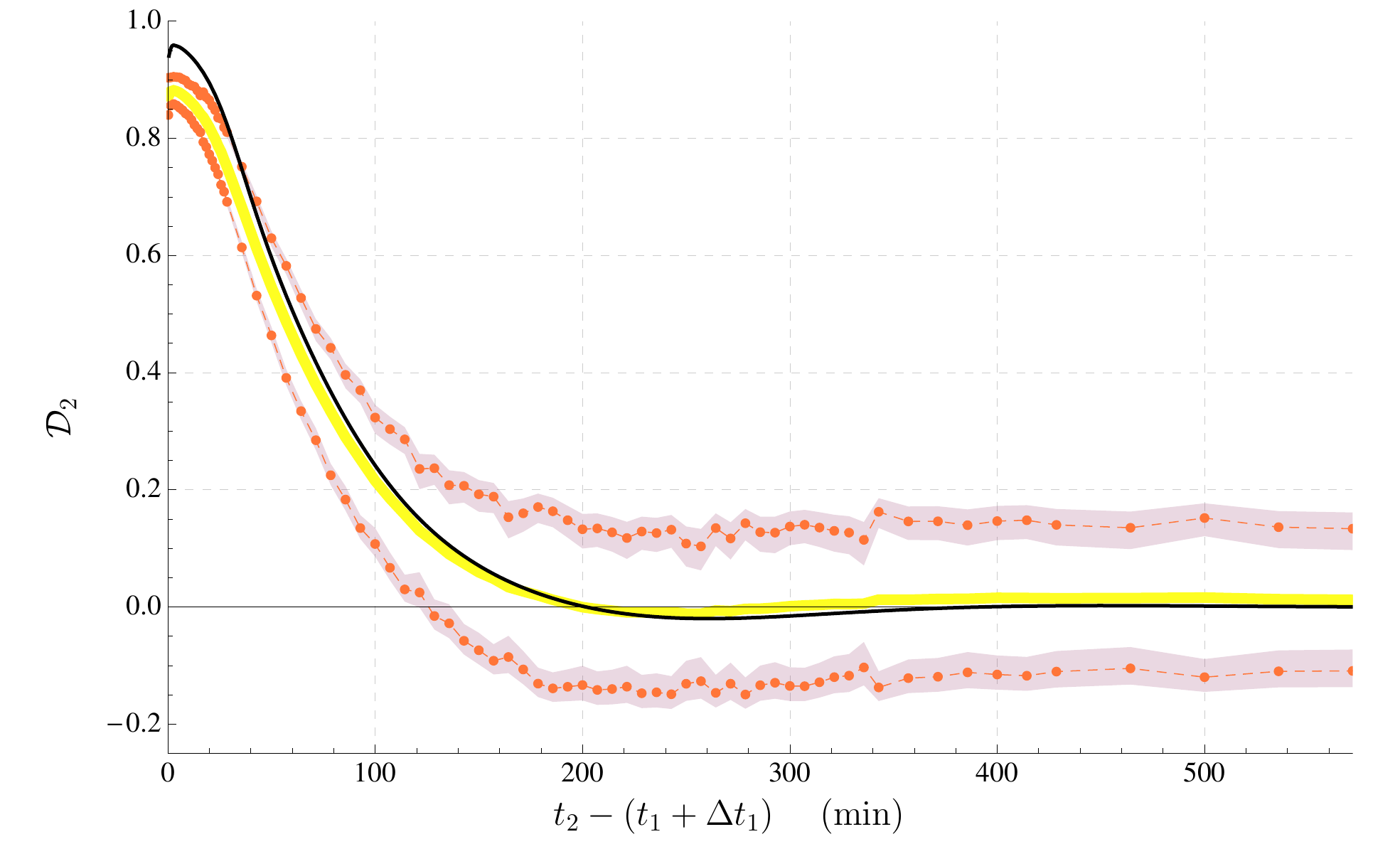}
\end{center}
\caption{%
}
\label{fig:hsr_desens_coeff}%
\end{figure}

Fig.~\ref{fig:hsr_desens_coeff} depicts value of the desensitisation
coefficient $\dc_{2}$ for the \tS\ species, with respect to the time gap
between heat-shocks. After the first heat-shock, at the time gap of the
approximated memory loss, i.e. at ca. 400\xu{\tu}, system is very close to the
homeostasis steady state (cf. Fig.~\ref{fig:hsr_nstab} in the Supporting
Material, \mbox{$t\approx \tgap_{1}+400\approx 470\xu{\tu}$}).

In the stochastic model we may observe a~non-zero (slightly positive)
level of mean $\dc_{2}$, after the thermotolerance effect has vanished.
More importantly, the stochastic variant presents a~constantly
high standard deviation of the desensitisation intensity: ca. 20\% of its
expected maximum level (which is observed for the very short time gap between
heat shocks).
These results, as well as the overall difference with respect to the
deterministic model, may be attributed to the stochastic noise and the fact
that we take a~maximum amount of \tS\ in Eq.~\ref{eq:response} to~measure its
toxic influence, not the mean value.

\subsection*{Hyperthermia in multimodal oncological strategies}


It has been hypothesised that because hyperthermia engages \abbr{HSR} mechanism
and because capacity of this mechanism is limited, especially in cancer cells,
hyperthermia enhances the toxicity induced by a~second modality of cancer
treatment~\citep{neznanov2011}.
This synergistic effect of hyperthermia and other cancer therapies can be
attributed to the much higher accumulation of denatured proteins (\tS), which
are deadly for cell. In our modelling approach we investigate, by means of
the presented mathematical \abbr{HSR} model, the temperature dependence of the
{\HSR} in combination with bortezomib-induced inhibition of proteasome.

In our intracellular-level model we assume that hyperthermia treatment
is represented by a~heat-shock with an immediate temperature shift, as presented
in previous section.
In order to incorporate into the model the inhibitory effect of bortezomib, we
limit the \abbr{HSP}-assisted degradation of denatured proteins
(Eq.~\ref{eq:r9}) and degradation of \abbr{HSP} itself
(Eq.~\ref{eq:r8}). More precisely, we linearly scale both reaction rate
constants \kr{k}{},
i.e., we set
$(1-\xI)\cdot\kr{k}{}$, for $\xI\in[0,1]$, where \xI\ represents current
inhibition level (when no drug is administrated $\xI=0$ whereas in case of
maximum inhibition $\xI=\xImax$).

We used bortezomib pharmacodynamics as modelled by Sung \&
Simon~\citep{sung2004}.
Namely, the inhibition level linearly raises up to its maximum level at
\tImax=60\xu{\tu}, after which it decays with
a~half-life $\tIhl=12\cdot\tImax$, i.e.:
\begin{equation}\label{eq:inh_curve}
\fAt{\xI}=\xImax\cdot
\begin{cases}
 \frac{\symTime}{\tImax} & \text{for }\symTime\leq\tImax \\
 \e^{-\ki (\symTime-\tImax)} & \text{for } \symTime>\tImax
\end{cases},
\end{equation}
where $\ki=\hfrac{\fofx{\ln}{2}}{\tIhl}$.
The maximum inhibition level \xImax\ directly corresponds to the drug dose.
For a~maximum tolerated bortezomib dose, \xImax\ is equal to ca. 65\%, while for
some of the next-generation proteasome inhibitors, such as carfilzomib or
ONX-0912, both of which are in clinical development, it was possible to reach
over 80\% proteasome inhibition in blood (with consecutive-day dosing
schedules)~\citep{molineaux2012}.

\begin{sidewaysfigure}
\begin{center}
\includegraphics[width=0.99\linewidth]{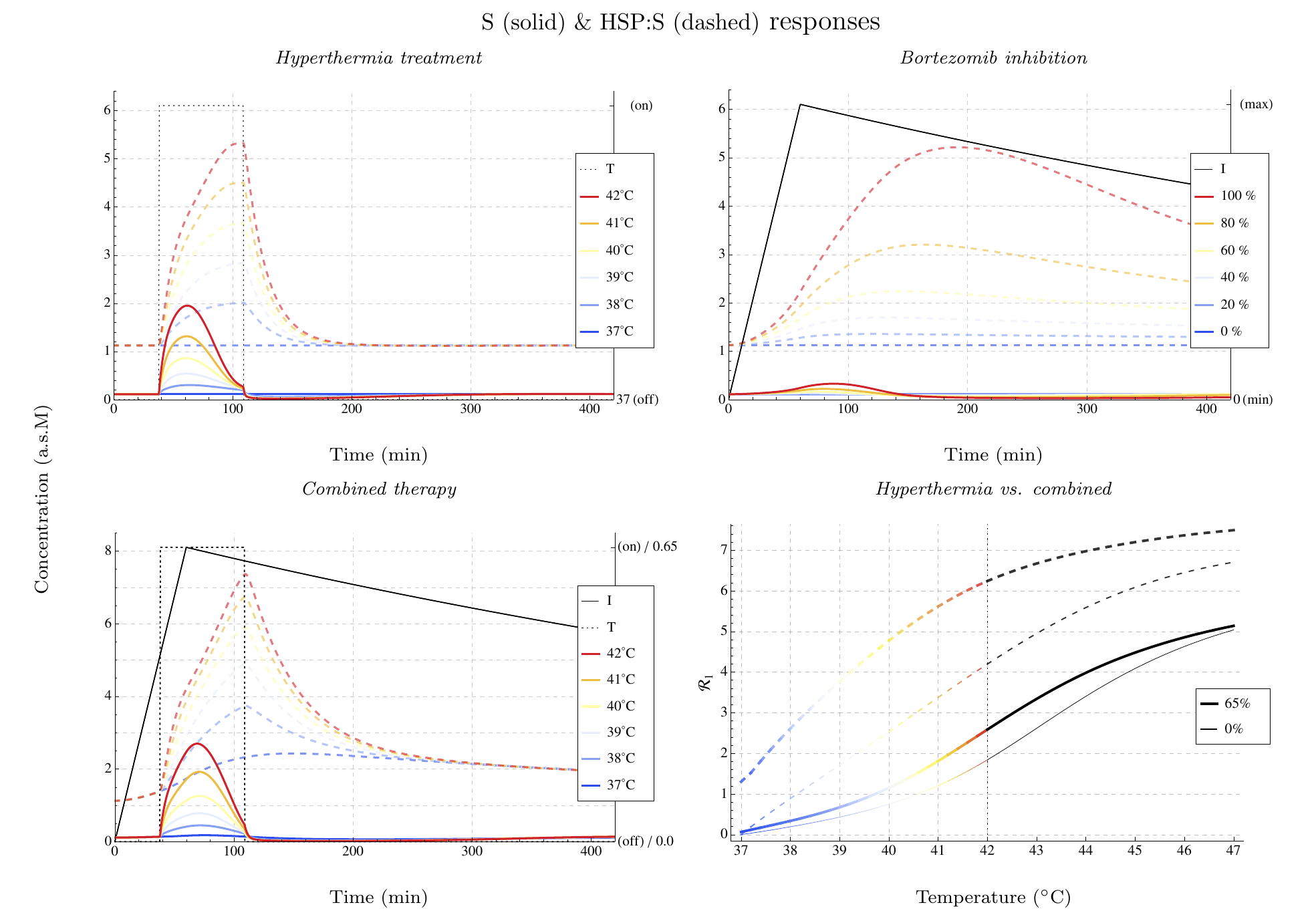}
\end{center}
\caption{%
}
\label{fig:hsr_sm_vs_mm_resp}%
\end{sidewaysfigure}

Fig.~\ref{fig:hsr_sm_vs_mm_resp} depicts activity of \tS\ and \thsps\ complex,
with respect to an~unimodal proteasome inhibition treatment for a~range of its
maximum levels \xImax, as well as a~unimodal hyperthermia treatment and
combined 65\% maximum inhibition treatment for a~range of moderate hyperthermia
temperatures. Recall that activity peak of a~cytotoxic \tS\ defines level of
{\HSR}~$\resp_1$ (Eq.~\ref{eq:response}). The higher the response is the more
effective is the therapy in terms of indicating a~higher probability of death of
a~cancer cell.

The bortezomib-based proteasome inhibition and hyperthermia induce a~very
similar total number of denatured proteins (see
Fig.~\ref{fig:hsr_sm_vs_mm_resp}). However, in case of proteasome
inhibition vast majority of these proteins is being secured in \thsps\ complexes
on the fly. This is due to the gradual increase of bortezomib inhibition effect,
which is not fast enough with respect to a~rate at which new \thsp\ molecules
are synthesised.
The immediate heating has a~much better effect in terms of \tS\ cytotoxicity.
Furthermore, when both therapies are applied simultaneously levels of both
\tS\ and \thsps\ complex indeed are higher than in case of an application of
only one of the treatment modalities. \abbr{HSR} capacity, as represented by an
analogous $\resp_1$ coefficient for \thsps\ complex (cf.
Eq.~\ref{eq:response}), is much closer to saturation plateau in case of the
65\% peak inhibition level than without inhibition (see
Fig.~\ref{fig:hsr_sm_vs_mm_resp}). Hence, increase of the temperature has
a~better effect in the combined treatment, in the sense of a~deadly accumulation
of free \tS\ molecules.

\begin{sidewaysfigure}
\begin{center}
    \includegraphics[width=0.99\linewidth]{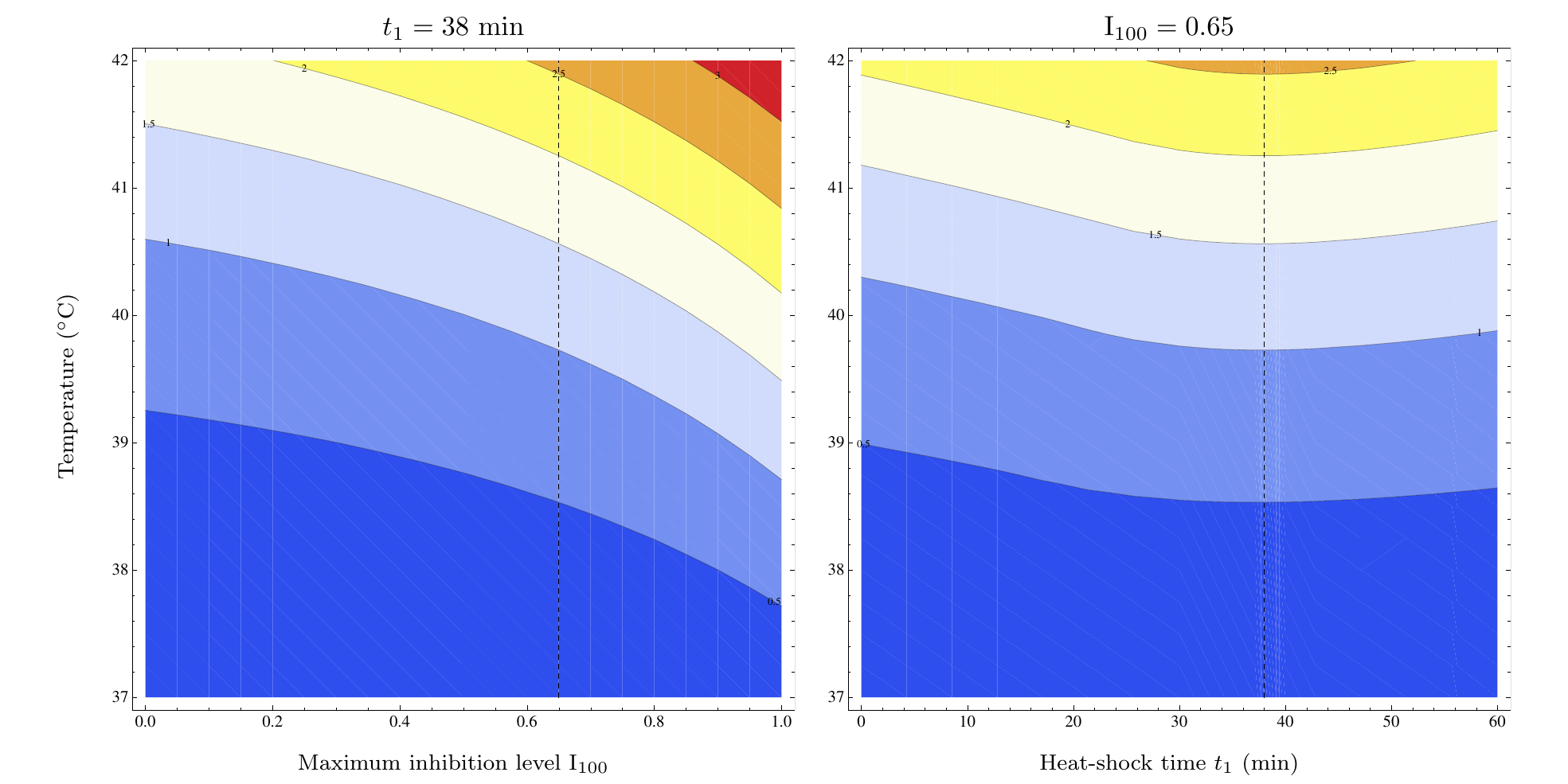}
\end{center}
\caption{%
}
\label{fig:hsr_mm_resp}%
\end{sidewaysfigure}

Fig.~\ref{fig:hsr_mm_resp} depicts this synergistic effect in a~continuous
scale of both the temperature and the maximum inhibition level of bortezomib.
A~monotone increase in response with respect to both modalities can be observed
regardless of the heat-shock application time (see
Fig.~\ref{fig:hsr_mm_resp_all} in the Supporting Material).
We~found that multimodal toxicity response increases by over 40\% with respect
to a~unimodal hyperthermia response
for a~maximum inhibition level equal to reported 65\%, up to over 80\% increase
for a~theoretical maximum of 100\% of proteasome inhibition.
Moreover, we established  $\symTime_1^{*}\approx 38\xu{\tu}$ as an~optimal time
to start hyperthermia treatment in combination with 65\% bortezomib inhibition
(see Fig.~\ref{fig:hsr_mm_resp}).
Interestingly, this is not in agreement with a~maximum area under the bortezomib
inhibition curve (AUC), a~common pharmacokinetic efficacy measure.
For the heat-shock duration $\tgap_1=71\xu{\tu}$, AUC maximum is reached
at $\symTime_1\approx 56\xu{\tu}$ (see Fig.~\ref{fig:hsr_inh_auc} in the
Supporting Material).
Timing of heat-shock in the optimal multimodal treatment strategy
$\symTime_1^{*}$ can be intuitively explained by the following observations
(cf. Fig.~\ref{fig:hsr_sm_vs_mm_resp}). Firstly, time required for denatured
proteins to peak after the beginning of a~heat-shock is roughly the same as the
time gap between $\symTime_1^{*}$ and \tImax\ (22\xu{\tu}). Secondly, at
$\symTime_1^{*}$ the inhibition itself has still a~relatively low impact.
This way, inhibition peak coincides with the period of maximum
temperature-induced toxicity, at which \abbr{HSR} mechanism is the most
occupied, thus effecting in the optimal synergistic toxicity.

\section*{Conclusions  and Discussion}%
\indent

We formalised and quantified the notion of thermotolerance induced by the
\abbr{HSP}-based mechanism of {\HSR}. Although we found a~deterministic approach
to be a~valid approximation of the stochastic \abbr{HSR} model, the latter
variant presented a~high level of intrinsic noise. In~consequence, we observe
a~significant level of intrinsic thermotolerance intensity
which can be highly increased by the heat-shock 
accompanied by a~high reduction of variability.
We would like to emphasise that in this analysis we demonstrated feasibility and
practical potential of the {\PMC} technique, more specifically its lesser
known approximate variant.

Next, by mathematical modelling of \abbr{HSR} we were able to support the common
belief that the combined cancer treatment strategies can more effectively
increase cytotoxicity of denatured proteins in cancer cells than unimodal
strategies. Moreover, we presented an optimal starting time for a~moderate
hyperthermia treatment in combination with a~proteasome inhibitor application.
This is an example of how mechanistic modelling can surpass pharmakokinetic
measures of optimal drug efficacy, such as AUC (which basically is an
optimisation only with respect to system's input).

We suggest that the synergistic effect of hyperthermia and other cancer
treatment modalities (like chemo- and radiotherapies) is caused by increased
accumulation of denatured proteins, i.e., heat and drug-sensitive proteins or
heat and radiation-sensitive proteins. This results in bigger demand for
heat-shock proteins and higher selective barrier for cells.

Our model-based analysis proves successful in reproducing experimental knowledge
of key aspects of hyperthermia treatment, and as such offers reasonable
framework for studying its connections with {\HSR}. However, all of the
kinetic models of molecular biological systems and means of their analysis are
incomplete due to constraints under which these models are formulated. In this
regard, we would like to point out that this is a~\poc\ model-based analysis and
there are many issues to address within this work.
For instance, we omitted the investigation of the day-based strategies of
multimodal treatment. This is because we found that in our \abbr{HSR} model the
single cell level thermotolerance duration (ca. 6.5h) is much shorter than the
bortezomib decay (12h half-life), thus, making the latter a~determining factor
for a~standard, daily dosing schedule. The inconsistency between reported
(24-48h) and simulated duration of thermotolerance can be primarily attributed
to the fact that thermotolerance, in general, is most likely not only induced by
the \abbr{HSP} upregulation \pcitep{cf.}{smith1991}.
Secondly, this inconsistency may also be attributed to the simplistic single
cell modelling of the immediate temperature shift, disregarding spatial heat
distribution and the preheating period as in, e.g., the whole-body hyperthermia
\pcitep{cf.}{wust2002}.
In this regard, to provide solid, quantitative results, our model requires
more extensive calibration with respect to experimental data,
including the behaviour for varying temperatures
\pcitep{cf.}{rieger2005,szymanska2009}.
Nevertheless, undoubtedly, our analysis gives a~valuable mathematical
framework for model-based understanding of hyperthermia treatment strategies,
such as these combining hyperthermia with very promising therapeutic proteasome
inhibitors.



\section*{Methods}
Model was defined using the \tech{SBML-shorthand}
notation~\citep{gillespie2006}, and automatically generated in the
\tech{SBML} format~\citep{hucka2003}.
The \abbr{ODE} model was numerically solved using the \tech{MathSBML}
package of the \tech{Mathematica} software~\citep{shapiro2004}.
The corresponding stochastic version of this model, represented by
\abbr{CME} or, equivalently, \abbr{CTMC} (cf.~\citep{charzynska2012}), was
analysed using the \abbr{PMC} technique. To ensure the feasibility of this
approach we have used approximate
\abbr{PMC} techniques (\abbr{APMC}), implemented recently in the \tech{PRISM}
tool \citep{kwiatkowska2011}. Consequently, all stochastic simulations and the
confidence interval-based \abbr{APMC} were done using \tech{PRISM}.
To create the \tech{PRISM} model, we used a~prototype \tech{SBML}
translator which generates model specification in the \tech{PRISM} language.
Minor adjustments, such as factorisation of parameters or accounting for
mass conservation laws
were done manually.

For means of modelling frameworks comparison and stochastic noise
quantification (see Text~\ref{txt:hsr_stoch} in the Supporting Material, Sec.~1)
as well as for thermotolerance quantification (see
Fig.~\ref{fig:hsr_desens_coeff}) we used \tech{PRISM} rewards to describe first
and second moments of, respectively, species variables as well as one minus
desensitisation coefficient (see Eq.~\ref{eq:desens_coeff}).
Text~\ref{txt:hsr_stoch} in the Supporting Material, Sec.~2 describes in
detail the unbiased estimators and their symmetric confidence intervals for
mean, variance, variance-to-mean ratio, and for standard deviation of both
species and desensitisation coefficient random variables.

To stay within \abbr{CTMC} framework and, consequently, to seamlessly perform
stochastic simulations or model checking, we introduced approximate stochastic
perturbation events based on $n$-counting Poisson processes. Precision of
a~single perturbation event, measured as a~standard deviation, is proportional
by square root to the number of counting levels~$n$ and inverse linearly
proportional to the expected time of occurrence of this event.
The approximate stochastic perturbation strategy for on and off heat-shock
events was encoded manually in \tech{PRISM} language, according to the scheme
presented in Text~\ref{txt:hsr_stoch} in the Supporting Material, Sec.~3.

\section*{Acknowledgements}
The work of ZS was supported by the Polish National Science Centre grant
2011/01/D/ST1/04133. The work of MR and SL was partially supported by the Polish
Ministry of Science and Higher Education grant N~N206 356036. The work of MR
and AG was partially supported by the Polish National Science Center grant
2011/01/B/NZ2/00864 and by the Biocentrum Ochota project
POIG.02.03.00-00-003/09.

MR carried out the \cs, prepared all figures and has written the manuscript.
ZS and SL prepared parts of the manuscript. ZS participated in model
adjustments. SL supervised the model checking experiments. AG supervised the
whole project and participated in drafting of the manuscript and improving of
the final manuscript.
All authors have read and approved the final manuscript
and there is no conflict of interest to declare.

All authors would like to thank to prof.~Maciej {\.Z}ylicz
(International Institute of Molecular and Cell Biology in Warsaw)
and prof. Bogdan Lesyng (University of Warsaw) for valuable discussions and for
inspiring this research.


\clearpage
\section*{Figure Legends}
\subsubsection*{Figure~\ref{fig:hsr_scheme}.}
Scheme of the \abbr{HSR} model. Squares represent species, including complexes,
and dots represent reactions, with substrates and products denoted respectively
by incoming and outgoing arrows.
On the left side of the scheme, the denaturation of {\tP}s \xP\ and refolding
or degradation of denatured proteins \xS\ (\tS) moderated by the \abbr{HSP}
chaperones. On the right side, the adaptive \abbr{HSP} production loop,
stimulated by \abbr{HSF}, which trimerise and initiate \abbr{HSE} transcription
and \tmrna\ translation (dotted arrow). As a~negative feedback, \abbr{HSP}
molecules promote \abbr{HSF} trimers dissociation and inhibit single \abbr{HSF}
molecules by direct binding.
The loop is closed by the inflowing \tS\ which forces out inhibited \abbr{HSF}
out of the complex with \abbr{HSP}.%

\subsubsection*{Figure~\ref{fig:hsr_activity}.}
Numerical simulations of the \abbr{HSR} \abbr{ODE} model for a~constant
42\cdeg\ heating strategy. Simulation starts at a~37\cdeg\ steady state.%
The upper plot depicts \abbr{HSP} response to the temperature-stimulated inflow
of denatured proteins \xS\ (\tS). Free \tS\ is instantaneously bound into
a~\xhsps\ complex. Insufficient amount of free \abbr{HSP} causes its extraction
from the \xhsphsf\ complex, forming an~initiative response of the cell.
Released in exchange \abbr{HSF} induces adaptive production of \abbr{HSP}
molecules to complement its deficiency as indicated by accumulation of \xS,
with peak at ca. 25\xu{\tu}. After over 120\xu{\tu} the excess of upregulated
\abbr{HSP} is used to inhibit \abbr{HSF} activity. System completely stabilises
after ca. 650\xu{\tu} (Fig.~\ref{fig:hsr_nstab} in the Supporting Material) with
most of constantly inflowing \xS\ secured in the \xhsps\ complexes. %
The lower plot depicts the adaptive \abbr{HSP} production, stimulated by
\abbr{HSF}. \abbr{HSF} trimerises and initiate \abbr{HSE} transcription
to \xmrna, followed by further translation to \abbr{HSP}, as visible by the
shifted activity of subsequent components.%

\subsubsection*{Figure~\ref{fig:hsr_desens}.}
Thermotolerance in the {\HSR}: the \tS\ activity
(solid) during the two consecutive immediate heat-shocks (dotted) of 5\cdeg\
over the homeostatis level of 37\cdeg. The intensity of intoxication resulting
from the amount of \tS\ (filled area) depends on the time gap between
heat-shocks.
Interestingly, activity of the \tS\ in the second shock can be
even higher than activity in the first shock, as shown for the time gap of
240\xu{\tu}. This is due to a~temporary deficit of \thsp\
(see.~Fig.~\ref{fig:hsr_desens_shift} in the Supporting Material for details).

\subsubsection*{Figure~\ref{fig:hsr_desens_coeff}.}
The desensitisation coefficient $\dc_{2}$ for the \tS\ in the
\abbr{ODE} model (black line) and its mean and standard deviation in
\abbr{CTMC}, plotted against the time gap between end of the first heat-shock
and the beginning of the second heat-shock. Duration of both
heat-shocks~$\tgap_n$ ($n=1,2$) is equal to 71\xu{\tu}.
Memory of the first heat-shock is lost when the desensitisation coefficient
value stabilises around 0, which is approximately at 400\xu{\tu} for both
mathematical models. %
Mean (yellow line) and standard deviation (orange line) of $\dc_{2}$ was
calculated at selected time points (dots).
Both estimators have a~confidence interval with 95\% confidence level.
In case of the mean value the confidence interval width is less than $5\cdot
10^{-3}$, whilst for the standard deviation the confidence interval is depicted
as a~strip.
Estimators were calculated using \abbr{APMC} with $10^{4}$ and $5\cdot 10^{4}$
independent simulation samples for the first and the second moment
respectively (see Text~\ref{txt:hsr_stoch} in the Supporting Material,
Sec.~2 for details).%

\subsubsection*{Figure~\ref{fig:hsr_sm_vs_mm_resp}.}
The {\HSR} with respect to different temperatures and to different inhibition
levels, applied separately (unimodal treatments) and simultaneously (combined
treatment).
The upper left plot depicts \abbr{ODE} trajectory of the \tS\ (solid lines) and
the \thsps\ complex (dashed lines), upon a~71\xu{\tu} heat-shock induced
at~38\xu{\tu} for $T=37,38\ldots,42\cdeg$. The upper right plot depicts
the same trajectories for bortezomib maximum inhibition levels
$\xImax=0,20,\ldots,100\%$.
In the similar manner the bottom left plot presents an~effect of combining both
therapies for varying $T$ and fixed $\xImax=65\%$.
Finally, the \tS\ toxic response coefficient $\resp_1$ (Eq.~\ref{eq:response})
and the analogous coefficient for \xhsps, measuring \abbr{HSR} capacity, are
depicted in the bottom right plot with respect to
$T\in\left[37,47\right]$, i.e. a~continuous temperature range, broadened for
a~context. For a~comparison $\resp_1$ coefficient curves are presented for both
a~unimodal hyperthermia treatment (thin lines) and a~treatment combined with
$\xImax=65\%$ (thick lines).

\subsubsection*{Figure~\ref{fig:hsr_mm_resp}.}
Contour plot of the {\HSR} level $\resp_1$ with respect to a~heat-shock
temperature (vertical axis) and with respect to a~maximum level of proteasome
inhibition for heat-shock applied at $\symTime_1=38\xu{\tu}$ (left plot), or
with respect to time of heat-shock application at maximum inhibition of
$\xImax=65\%$ (right plot). %
Heat-shock takes $\tgap_1=71\xu{\tu}$. %
Level of~$\resp_1$ (Eq.~\ref{eq:response}), denoted on the plot by colours
from blue (the weakest) to red (the strongest), measures the toxicity of the
combined therapy. %
Dashed vertical line at each plot denotes conditions for the other plot. Choice
for maximum inhibition level \xImax\ was driven by data reported in literature
(see text for details), whereas choice for heat-shock time was based on
maximisation of the multimodal strategy response (cf. right plot).

\clearpage
\section*{Figures}

\clearpage
\section*{Supporting Material index}

\renewcommand{\thefigure}{S\arabic{figure}}
\setcounter{figure}{0}

\begin{figure}[!h]
\caption{%
Stability of the HSR ODE model.
\hspace{\linewidth}%
}
\label{fig:hsr_nstab}%
\end{figure}

\begin{figure}[!h]
\caption{%
The \thseO\ fit to the experimental data.
\hspace{\linewidth}%
}
\label{fig:hsr_fit}%
\end{figure}

\renewcommand\figurename{Text}
\begin{figure}[!h]
\caption{%
Stochastic modelling of HSR
\hspace{\linewidth}%
}
\label{txt:hsr_stoch}%
\end{figure}
\renewcommand\figurename{Figure}

\begin{figure}[!h]
\caption{%
The \tS\ and  \thsp\ response to the two consecutive immediate heat-shocks.
\hspace{\linewidth}%
}
\label{fig:hsr_desens_shift}%
\end{figure}

\begin{figure}[!h]
\caption{%
Contour plots of the heat-shock response level for multiple equally-distributed
heat-shock application times.
\hspace{\linewidth}%
}
\label{fig:hsr_mm_resp_all}%
\end{figure}

\begin{figure}[!h]
\caption{%
Area under the bortezomib inhibition curve (AUC) versus heat-shock application
time.
\hspace{\linewidth}%
}
\label{fig:hsr_inh_auc}%
\end{figure}

\section*{Additional Files index}

\renewcommand{\thefigure}{F\arabic{figure}}
\setcounter{figure}{0}
\renewcommand\figurename{File}
\begin{figure}[!h]
\caption{%
XML file in the SBML format containing deterministic model of HSR.
\hspace{\linewidth}%
}
\label{fil:sbml}%
\end{figure}
\begin{figure}[!h]
\caption{%
Text file in the PRISM model format containing stochastic model of HSR.
\hspace{\linewidth}%
}
\label{fil:prism}%
\end{figure}
\renewcommand\figurename{Figure}


\begin{thebibliography}{10}

\bibitem{mayer2005}
Mayer, M. P. \& Bukau, B.
\newblock 2005
\newblock Hsp70 chaperones: cellular functions and molecular mechanism.
\newblock \textit{Cell. Mol. Life Sci.} \textbf{62}, 670--684.

\bibitem{georgopoulos1993}
Georgopoulos, C. \& Welch, W. J.
\newblock 1993
\newblock Role of the major heat shock proteins as molecular chaperones.
\newblock \textit{Ann. Rev. Cell Biol.} \textbf{9}, 601--634.

\bibitem{parsell1993}
Parsell, D. A. \& Lindquist, S.
\newblock 1993
\newblock The function of heat-shock proteins in stress tolerance: degradation
  and reactivation of damaged proteins.
\newblock \textit{Ann. Rev. Genet.} \textbf{27}, 437--496.

\bibitem{barnes2001}
Barnes, J. A., Dix, D. J., Collins, B. W., Luft, C. \& Allen, J. W.
\newblock 2001
\newblock Expression of inducible {Hsp70} enhances the proliferation of {MCF-7}
  breast cancer cells and protects against the cytotoxic effects of
  hyperthermia.
\newblock \textit{Cell Stress Chaperon.} \textbf{6}, 316--325.

\bibitem{wust2002}
Wust, P., Hildebrandt, B., Sreenivasa, G., Rau, B., Gellermann, J., Riess, H.,
Felix, R. \& Schlag, P. M.
\newblock 2002
\newblock Hyperthermia in combined treatment of cancer.
\newblock \textit{Lancet. Oncol.} \textbf{3}, 487--497.

\bibitem{hildebrandt2002}
Hildebrandt, B., Wust, P., Ahlers, O., Dieing, A., Sreenivasa, G., Kerner, T.,
Felix, R. \& Riess, H.
\newblock 2002
\newblock The cellular and molecular basis of hyperthermia.
\newblock \textit{Crit. Rev. Oncol. Hematol.} \textbf{43}, 33--56.

\bibitem{van_der_zee2002}
van~der Zee, J.
\newblock 2002
\newblock Heating the patient: a promising approach?
\newblock \textit{Ann. Oncol.} \textbf{13}, 1173--1184.

\bibitem{neznanov2011}
Neznanov, N., Komarov AP, Neznanova, L., {Stanhope-Baker}, P. \& Gudkov, A. V.
\newblock 2011
\newblock Proteotoxic stress targeted therapy {(PSTT):} induction of protein
  misfolding enhances the antitumor effect of the proteasome inhibitor
  bortezomib.
\newblock \textit{Oncotarget} \textbf{2}, 209--221.

\bibitem{molineaux2012}
Molineaux, S. M.
\newblock 2012
\newblock Molecular pathways: targeting proteasomal protein degradation in
  cancer.
\newblock \textit{Clin. Cancer Res.} \textbf{18}, 15--20.

\bibitem{szymanska2009}
Szyma{\'n}ska, Z. \& {\. Z}ylicz, M.
\newblock 2009
\newblock Mathematical modeling of heat shock protein synthesis in response to
  temperature change.
\newblock \textit{J. Theor. Biol.} \textbf{259}, 562--569.

\bibitem{peper1998}
Peper, A., Grimbergen, C. A., Spaan, J. A., Souren, J. E. \& van Wijk, R.
\newblock 1998
\newblock A mathematical model of the hsp70 regulation in the cell.
\newblock \textit{Int. J. Hyperther.} \textbf{14}, 97--124.

\bibitem{petre2011}
Petre, I., Mizera, A., Hyder, C. L., Meinander, A., Mikhailov, A.,
Morimoto, R. I., Sistonen, L., Eriksson, J. E. \& Back, R.
\newblock 2011
\newblock A simple mass-action model for the eukaryotic heat shock response and
  its mathematical validation.
\newblock \textit{Nat. Comp.} \textbf{10}, 595--612.

\bibitem{mizera2010}
Mizera, A. \& Gambin B.
\newblock 2010
\newblock Stochastic modelling of the eukaryotic heat shock response.
\newblock \textit{J. Theor. Biol.} \textbf{265}, 455--466.

\bibitem{lepock1993}
Lepock, J. R., Frey, H. E. \& Ritchie, K. P.
\newblock 1993
\newblock Protein denaturation in intact hepatocytes and isolated cellular
  organelles during heat shock.
\newblock \textit{J. Cell Biol.} \textbf{122}, 1267--1276.

\bibitem{abravaya1991}
Abravaya, K., Phillips, B. \& Morimoto, R. I.
\newblock 1991
\newblock Attenuation of the heat shock response in {HeLa} cells is mediated by
  the release of bound heat shock transcription factor and is modulated by
  changes in growth and in heat shock temperatures.
\newblock \textit{Genes Dev.} \textbf{5}, 2117--2127.

\bibitem{kwiatkowska2011}
Kwiatkowska, M., Norman, G. \& Parker, D.
\newblock 2011
\newblock {PRISM} 4.0: verification of probabilistic real-time systems.
\newblock \textit{Lect. Notes. Comput. Sc.} \textbf{6806}, 585--591.

\bibitem{hucka2003}
Hucka, M., Finney, A., Sauro, H. M., Bolouri, H., Doyle, J. C., Kitano, H.,
Arkin, A. P., Bornstein, B. J., Bray, D., Cornish-Bowden, A. et~al.
\newblock 2003
\newblock The systems biology markup language (SBML): a medium for
  representation and exchange of biochemical network models.
\newblock \textit{Bioinformatics} \textbf{19}, 524--531.

\bibitem{sung2004}
Sung, M. H. \& Simon, R.
\newblock 2004
\newblock In silico simulation of inhibitor drug effects on nuclear
  factor-{kappaB} pathway dynamics.
\newblock \textit{Mol. Pharmacol.} \textbf{66}, 70--75.

\bibitem{smith1991}
Smith, B. J. \& Yaffe, M. P.
\newblock 1991
\newblock Uncoupling thermotolerance from the induction of heat shock proteins.
\newblock \textit{Proc. Natl. Acad. Sci. USA} \textbf{88}, 11091--11094.

\bibitem{rieger2005}
Rieger, T. R., Morimoto, R. I. \& Hatzimanikatis, V.
\newblock 2005
\newblock Mathematical modeling of the eukaryotic heat-shock response: dynamics
  of the hsp70 promoter.
\newblock \textit{Biophys. J.} \textbf{88}, 1646--1658.

\bibitem{gillespie2006}
Gillespie, C. S., Wilkinson, D. J., Proctor, C. J., Shanley, D. P., Boys, R. J.
\& Kirkwood, T. B. L.
\newblock 2006
\newblock Tools for the {SBML}. Community.
\newblock \textit{Bioinformatics} \textbf{22}, 628--629.

\bibitem{shapiro2004}
Shapiro, B. E., Hucka, M., Finney, A. \& Doyle, J.
\newblock 2004
\newblock {MathSBML:} a package for manipulating {SBML-based} biological
  models.
\newblock \textit{Bioinformatics} \textbf{20}, 2829--2831.

\bibitem{charzynska2012}
Charzy{\'n}ska, A., Na{\l}ecz, A., Rybi{\'n}ski, M. \& Gambin, A.
\newblock 2012
\newblock Sensitivity analysis of mathematical models of signaling pathways.
\newblock \textit{BioTechnol.} \textbf{93}, 291--308.

%

\end{thebibliography}
\end{document}